\documentclass[preprint,prd,nofootinbib]{revtex4}
\usepackage[dvips]{graphicx}
\usepackage{subfigure}
\usepackage{latexsym}

\newcommand{\nc}{\newcommand}
\nc{\beq}{\begin{equation}}
\nc{\eeq}{\end{equation}}
\nc{\beqa}{\begin{eqnarray}}
\nc{\eeqa}{\end{eqnarray}}
\nc{\bert}{\raise-.55mm\hbox{\large$\Box$}} 

 \def\lsim{\mathrel{\rlap{\lower4pt\hbox{\hskip1pt$\sim$}}
\raise1pt\hbox{$<$}}}

\def\gsim{\mathrel{\rlap{\lower4pt\hbox{\small \hskip1pt$\sim$}}
 \raise1pt\hbox{$>$}}}

\def\lsim{\mathrel{\rlap{\lower4pt\hbox{\small \hskip1pt$\sim$}}
 \raise1pt\hbox{$<$}}}
 
\newwrite\ffile\global\newcount\figno \global\figno=1

\def\writedef#1{}
\def\figin{\epsfcheck\figin}\def\figins{\epsfcheck\figins}

\def\epsfcheck{\ifx\epsfbox\UnDeFiNeD \message{(NO epsf.tex, FIGURES
WILL BE IGNORED)}
\gdef\figin##1{\vskip2in}\gdef\figins##1{\hskip.5in}
instead \else\message{(FIGURES WILL BE INCLUDED)}%
\gdef\figin##1{##1}\gdef\figins##1{##1}\fi} \def\figinsert{}
\def\ifig#1#2#3{\xdef#1{fig.~\the\figno} \writedef{#1\leftbracket
fig.\noexpand~\the\figno}%
\figinsert\figin{\centerline{#3}}\medskip\centerline{\vbox{\baselineskip12pt
\advance\hsize by -1truein\center\footnotesize{ Fig.~\the\figno.} #2}}
\bigskip\endinsert\global\advance\figno by1}
\def\endinsert{}

\begin{document}

\title{~\\ Spontaneous Lorentz Violation and the Long-Range
Gravitational Preferred-Frame Effect}

\author{Michael~L.~Graesser\footnote{graesser@theory.caltech.edu},
Alejandro~Jenkins\footnote{jenkins@theory.caltech.edu}, and
Mark~B.~Wise\footnote{wise@theory.caltech.edu}}

\affiliation{California Institute of Technology, Pasadena, CA 91125
\bigskip \bigskip \bigskip \bigskip}

\preprint{hep-th/0501223}
\preprint{CALT-68-2538}

\begin{abstract}

\bigskip

Lorentz-violating operators involving Standard Model fields are
tightly constrained by experimental data. However, bounds are more
model-independent for Lorentz violation appearing in purely
gravitational couplings. The spontaneous breaking of Lorentz
invariance by the vacuum expectation value of a vector field selects a
universal rest frame. This affects the propagation of the graviton,
leading to a modification of Newton's law of gravity. We compute the
size of the long-range preferred-frame effect in terms of the
coefficients of the two-derivative operators in the low-energy
effective theory that involves only the graviton and the Goldstone
bosons.

\end{abstract}

\maketitle


\newpage

The possibility of breaking Lorentz invariance is of interest to
theoretical physicists for a variety of reasons. For instance,
spontaneous Lorentz violation has been proposed at various times as a
starting point in alternative theories of electrodynamics
\cite{diracbjorken} and of linear gravity \cite{lgravity}, and as a
possible solution to the horizon problem in cosmology \cite{moffat}.
Lorentz violation has also been discussed as a potentially observable
signal of physics beyond the Planck scale (whether in the context of
string theory, noncommutative geometry, or loop quantum gravity)
\cite{kosteleckysamuel}, and some researchers have claimed that there
is evidence of such violation in the measurements of the energies of
cosmic rays \cite{colemanglashow}.

Experimental data puts very tight constraints on Lorentz violating
operators that involve Standard Model particles \cite{kostelecky}, but
the bounds are more model-independent on Lorentz violation that
appears only in couplings to gravity \cite{nelson,burgess}. One broad
class of Lorentz-breaking gravitational theories are the so-called
vector-tensor theories in which the space-time metric $g^{\mu\nu}$ is
coupled to a vector field $S^\mu$ which does not vanish in the
vacuum. Consideration of such theories dates back to \cite{nordtvedt}
and their potentially observable consequences are extensively
discussed in \cite{Will}. These theories have an unconstrained
vector-field coupled to gravity. Theories with a unit constraint on
the vector field were proposed as a means of alleviating the
difficulties that plagued the original unconstrained theories
\cite{jacobson0}.

The phenomenology of these theories with the unit constraint has has
been recently explored. It has been proposed as a toy model for
modifying dispersion relations at high energy \cite{modifieddisp}. The
spectrum of long-wavelength excitations is discussed in
\cite{jacobson}, where it was found that all polarizations have a
relativistic dispersion relation, but travel with different
velocities. Applications of these theories to cosmology have been
considered in \cite{carroll,lim}. Constraints on these theories are
weak, as for instance, there are no corrections to the Post-Newtonian
parameters $\gamma$ and $\beta$ \cite{jacobson2}. The status of this
class of theories, also known as `Aether-theories', is reviewed in
\cite{unitvtreview}.

Here we begin by considering the general low-energy effective action
for a theory in which Lorentz invariance is spontaneously broken by
the vacuum expectation value (vev) of a Lorentz four-vector
$S^\mu$. With an appropriate rescaling, the vev satisfies \beq \langle
S_\mu S^\mu \rangle = 1 ~, \label{vev} \eeq since we assume the vev of
$S^{\mu}$ is time-like. The existence of this vev implies that there
exists a universal rest frame (which we sometimes refer to as the
`preferred frame') in which $S^{\mu} =\delta^{\mu}_0$. When the
resulting low-energy effective action is minimally coupled to gravity,
we shall see that it simply becomes the vector-tensor theory with the
unit constraint.

Objects of mass $M_1$ and $M_2$ in a system moving relative to the
preferred-frame can experience a modification to Newton's law of
gravity of the form \cite{Will,Will2} \beq U_{\rm Newton} = -G_{\rm N}
\frac{M_1 M_2}{r} \left(1- \frac{\alpha_2}{2} \frac{(\vec{w}\cdot
\vec{r})^2}{r^2}\right) \label{newtonlaw} \eeq where $\vec{w}$ is the
velocity of the system under consideration, such as the solar-system
or Milky Way galaxy, relative to the universal rest frame. The main
purpose of this note is to compute $\alpha_2$ in theories where
Lorentz invariance is spontaneously broken by the vev of a
four-vector. This PPN coefficient is more strongly constrained by
experiment than the other PPN parameters $\gamma$ and $\beta$
\cite{Will2}, so it is natural to focus on it.

The vev of $S^{\mu}$ spontaneously breaks Lorentz invariance. But as
rotational invariance is preserved in the preferred frame, only the
three boost generators of the Lorentz symmetry are spontaneously
broken. The low-energy fluctuations $ S^{\mu}(x)$ which preserve
Eq. (\ref{vev}) are the Goldstone bosons of this breaking, i.e., those
that satisfy \beq S_{\mu}(x) S^{\mu}(x) = 1 ~. \label{constraint} \eeq
In the preferred-frame the fluctuations can be parameterized as a
local Lorentz transformation \beq S^{\mu}(x) = \Lambda^{\mu}_0(x)=
\frac{1}{\sqrt{1-\vec{\phi}^2}} \left(\matrix{1 \cr
\vec{\phi}\cr}\right)~. \eeq

Under Lorentz transformations $S^{\mu}(x) \rightarrow
\Lambda^{\mu}_{\nu} S^{\nu}(x)$ and the symmetry is realized
non-linearly on the fields $\phi^{i}$. Using this field $S^{\mu}(x)$
we may then couple the Goldstone bosons to Standard Model
fields. Since however, the constraints on Lorentz-violating operators
\footnote{More correctly, operators that appear to be Lorentz
violating when the Goldstone bosons $\phi^{i}$ are set to zero.}
involving Standard Model fields are considerable \cite{kostelecky}, we
instead focus on their couplings to gravity, which are more model
independent because they are always present once the Goldstone bosons
are made dynamical.

The Goldstone bosons are made dynamical by adding in kinetic terms for
them. Since Lorentz invariance is only broken spontaneously, the
action for the kinetic terms should still be invariant under Lorentz
transformations. The only interactions relevant at the two
derivative-level and not eliminated by the constraint Eq.
(\ref{constraint}) are \beq {\cal L} = c_1 \partial_{\alpha}S^{\beta}
\partial^{\alpha} S_{\beta} +(c_2+c_3) \partial_{\mu} S^{\mu} \partial
_{\nu} S^{\nu} + c_4 S^{\mu} \partial_{\mu} S^{\alpha} S^{\nu}
\partial_{\nu} S_{\alpha} ~. \eeq Expanding this action to quadratic
order in $\phi^i$, one finds that the four parameters $c_i$ can be
chosen to avoid the appearance of any ghosts. In particular, we
require $c_1+c_4<0$.

With gravity present the situation is more subtle. One expects the
gravitons to `eat' the Goldstone bosons, producing a more complicated
spectrum \cite{kosteleckygravity,gripaios}. The covariant
generalization of the constraint equation becomes \beq g_{\mu \nu} (x)
S^{\mu}(x) S^{\nu}(x) =1 \label{constraint2} \eeq and in the action
for $S^{\mu}$ we replace $\partial_{\mu} \rightarrow
\nabla_{\mu}$. Note that there is no ``Higgs mechanism'' to give the
graviton a mass, since the connection is linear in derivatives of the
metric.

Local diffeomorphisms can now be used to gauge away the Goldstone
bosons. For under a local diffeomorphism (which preserves the
constraint Eq. (\ref{constraint2})), \beq S^{\prime
\mu}(x^{\prime})=\frac{\partial x^{\prime \mu}}{\partial x^{\nu}}
S^{\nu}(x) \eeq and with $x^{\prime \mu}=x^{\mu } + \epsilon^{\mu}$,
$S^{\mu} \equiv v^{\mu} + \phi^{\mu}$, \beq \phi^{\prime
\mu}(x^{\prime})= \phi^{\mu}(x) +v^{\rho} \partial_{\rho} \epsilon
^{\mu} \eeq from which we can determine $\epsilon^\mu$ to completely
remove $\phi^{\mu}$. Note that in the preferred frame, $\epsilon^{i}$
can be used to remove $\phi^{i}$. In this gauge, the constraint
Eq. (\ref{constraint2}) reduces to \beq
S^0(x)=\left(1-h_{00}(x)/2\right)~.\eeq The residual gauge invariance
left in $\epsilon^0$ can be used to remove $h_{00}$. This is an
inconvenient choice when the sources are static. In a more general
frame with $\langle S^{\mu} \rangle =v^{\mu}$, obtained by a uniform
Lorentz-boost from the preferred-frame, the constraint
Eq. (\ref{constraint2}) is solved by \beq S^{\mu}(x) = v^{\mu}
\left(1- v^{\rho} v^{\sigma} h_{\rho \sigma}(x) /2 \right)
~. \label{generalconstraint} \eeq

Next we discuss a toy model that provides an example of a more
complete theory, that at low-energies reduces to the theory described
above with the vector field satisfying a unit covariant constraint
(\ref{constraint2}).\footnote{For a related example, see
\cite{gripaios}.} Consider the following non-gauge invariant theory
for a vector boson $A^{\mu}$, \beq {\cal L} =-\frac{1}{2} g_{\mu \nu}
g^{\rho \sigma} \nabla_{\rho} A^{\mu} \nabla_{\sigma} A^{\nu} +
\lambda \left( g_{\mu \nu} A^{\mu} A^{\nu} - v^2 \right)^2 ~. \eeq
Fluctuations about the minimum are given by \beq g_{\mu \nu} =
\eta_{\mu \nu} + h_{\mu \nu} ~~,~~ A^{\mu} = v^{\mu}
+\psi^{\mu}~. \eeq This theory has one massive state $\Phi$ with mass
$M_{\Phi} \propto \lambda^{1/2} v$, which is \beq \Phi= v^{\mu}
\psi_{\mu} + h_{\mu \nu}v^{\mu } v^{\nu}/2~. \eeq In the limit that
$\lambda \rightarrow \infty$ this state decouples from the remaining
massless states. In the preferred frame the only massless states are
$h_{\mu \nu}$, and $\psi^i$. Since we have decoupled the heavy state,
we should expand \beq A^0 = v+\left[ \psi^0 + v h_{00}/2 \right] - v
h_{00}/2 \rightarrow v - v h_{00}/2 \eeq where in the last limit we
have decoupled the heavy state. Note that this parameterization of
$A^0$ is precisely the same parameterization that we had above for
$S^0$. In other words, in the limit that we decouple the only heavy
state in this model, the field $A^{\mu}$ satisfies $g_{\mu \nu} A^{\mu
} A^{\nu}=v^2$, which is the same as the constraint
(\ref{constraint2}) with $A^{\mu} \rightarrow v S^{\mu}$.

In the unitary gauge with $\phi^i=0$, the only massless degrees of
freedom are the gravitons. There are the two helicity modes which in
the Lorentz-invariant limit correspond to the two spin-2 gravitons,
along with three more helicities that are the Goldstone bosons, for a
total of five. The sixth would-be helicity mode is gauged away by the
remaining residual gauge invariance.

But the model that we started from does have a ghost, since we wrote a
kinetic term for $A^{\mu}$ that does not correspond to the
conventional Maxwell kinetic action. The ghost in the theory is $A^0$,
which in our case is massive. The presence of this ghost means that
this field theory model is not a good high-energy completion for the
low-energy theory involving only $S^{\mu}$ and gravity which we are
considering in this letter. We assume that a sensible high energy
completion exists for generic values of the $c_i$'s.

Now we proceed to compute the preferred-frame coefficient $\alpha_2$
appearing in the modification to Newton's law.

The action we consider is \beq S= \int d^4 x \, \sqrt{g} \left( {\cal
 L}_{\rm EH} + {\cal L}_{\rm V} + {\cal L}_{\rm gf} \right)
 \label{aether} \eeq with\footnote{The coefficients $c_i$ appearing
 here are related to those appearing in, for example \cite{jacobson},
 by $c_i^{\rm here}=-c^{\rm there}_i/16 \pi G$.} \beq {\cal L}_{\rm
 EH} = -\frac{1}{16 \pi G} R \eeq \beqa {\cal L}_{V}&=& c_1
 \nabla_{\alpha}S^{\beta} \nabla^{\alpha} S_{\beta} +c_2 \nabla_{\mu}
 S^{\mu} \nabla _{\nu} S^{\nu} + c_3 \nabla _{\mu }S^{\nu} \nabla
 _{\nu} S^{\mu} + c_4 S^{\mu} \nabla_{\mu} S^{\alpha} S^{\nu}
 \nabla_{\nu} S_{\alpha} ~, \eeqa and we use the metric signature
 $(+---)$. This is the most general action involving two derivatives
 acting on $S^{\mu}$ that contributes to the two-point function. Note
 that a coefficient $c_3$ appears, since in curved spacetime covariant
 derivatives do not commute. Other terms involving two derivatives
 acting on $S^{\mu}$ may be added to the action, but they are either
 equivalent to a combination of the operators already present (such as
 adding $R_{\mu \nu}S^{\mu}S^{ \nu})$, or they vanish because of the
 constraint Eq. (\ref{constraint2}). We assume generic values for the
 coefficients $c_i$ that in the low energy effective theory give no
 ghosts or gradient instabilities.

As previously discussed, $S^{\mu}$ satisfies the constraint
(\ref{constraint2}). We also assume that it does not directly couple
to Standard Model fields. In the literature, Eq. (\ref{constraint2})
is enforced by introducing a Lagrange-multiplier into the action. Here
we enforce the constraint by directly solving for $S^{\mu}$, as given
by Eq. (\ref{generalconstraint}), and then insert that solution back
into the action to obtain an effective action for the metric.

In our approach there is a residual gauge invariance which in the
preferred-frame corresponds to reparameterizations involving
$\epsilon^0$ only. To completely fix the gauge we add the gauge-fixing
term \beq {\cal L}_{\rm gf}= -\frac{\alpha }{2} \left(S^{\rho}
S^{\sigma} S^{\mu} \partial _{\mu} h_{\rho \sigma} \right)^2 ~. \eeq
Neglecting interaction terms, in the preferred frame the gauge-fixing
term reduces to \beq {\cal L}_{\rm gf}= -\frac{\alpha }{2}\left(
\partial _0 h_{00}\right)^2 ~. \eeq Physically, this corresponds in
the $\alpha \rightarrow \infty$ limit to removing all time-dependence
in $h_{00}$, without removing the static part which is the
gravitational potential. This is a convenient gauge in which to
compute when the sources are static.

At the two-derivative level, the only effect in this gauge of the new
operators is to modify the kinetic terms for the graviton. The
dispersion relation for the five helicities will be of the
relativistic form $E = \beta |\vec{k}|$, but where the velocities
$\beta$ are not the same for all helicities and depend on the
parameters $c_i$ \cite{jacobson}. This spectrum is different than that
which is found in the `ghost condensate' theory, where in addition to
the two massless graviton helicities, there exists a massless scalar
degree-of-freedom with a non-relativistic dispersion relation
\cite{nima}.

There exists a range for the $c_i$'s in which the theory has no ghosts
and no gradient instabilities \cite{jacobson}. In particular, for
small $c_i$'s, no gradient instabilities appear if \beq
\frac{c_1+c_2+c_3}{c_1+c_4} > 0 ~~~~\mbox{and}~~~~ \frac{c_1}{c_1+c_4}
> 0~. \eeq The condition for having no ghosts is simply $c_1+c_4 < 0$.

The correction to Newton's law in Eq. (\ref{newtonlaw}) is linear
order in the source. Thus to determine its size we only need to find
the graviton propagator, since the non-linearity of gravity
contributes at higher order in the source. In order to compute that
term we have to specify a coordinate system, of which there are two
natural choices. In the universal rest frame the sources, such as the
solar system or Milky Way galaxy, will be moving and the computation
is involved. We instead choose to compute in the rest frame of the
source, which is moving at a speed $|\vec{w}| \ll 1$ relative to the
universal rest frame. Observers in that frame will observe the Lorentz
breaking vev $v^{\mu} \simeq (1, -\vec{w})$. In the rest frame of the
source, a modified gravitational potential will be
generated. Technically this is because terms in the graviton
propagator $v \cdot k \simeq \vec{w} \cdot \vec{k}$ are
non-vanishing. It is natural to assume that dynamical effects align
the universal rest frame where $v^{\mu} = \delta^{\mu}_0$ with the
rest frame of the cosmic microwave background.

In a general coordinate system moving at a constant speed with respect
to the universal frame the Lorentz-breaking vev will be a general
time-like vector $v^{\mu}$. Thus we need to determine the graviton
propagator for a general time-like constant $v^{\mu}$. Since Lorentz
invariance is spontaneously broken, the numerator of the graviton
propagator is the most general tensor constructed out of the vectors
$v^{\mu}$, $k^{\nu}$ and the tensor $\eta^{\rho \sigma}$. There are 14
such tensors. Writing the action for the gravitons as \beq S=
\frac{1}{2} \int d^4 k \, \tilde{h}^{\alpha \beta}(-k) K_{\alpha \beta
| \sigma \rho}(k) \tilde{h}^{\sigma \rho}(k) \eeq it is a
straightforward exercise to determine the graviton propagator ${\cal
P}$ by solving \beq K_{\alpha \beta | \mu \nu}(k) {\cal P}^{\mu \nu |
\rho \sigma}(k) =\frac{1}{2} \left( \eta_\alpha^\rho \eta_\beta^\sigma
+ \eta_\alpha^\sigma \eta_\beta^\rho \right) ~.\eeq The above set of
conditions leads to 21 linear equations which determine the 14
coefficients of the graviton propagator in terms of the coefficients
$c_i$ and the vev $v^{\mu}$. Seven equations are redundant and provide
a non-trivial consistency check on our calculation.

Although it is necessary to compute all 14 coefficients in order to
invert the propagator, here we present only those which modify
Newton's law as described previously (assuming stress-tensors are
conserved for sources). These are \beqa {\cal P}^{\alpha \beta | \rho
\sigma} _{\rm Newton} &=& \left\{{\rm A} \eta^{\alpha \beta}
\eta^{\rho \sigma} + {\rm B }(\eta^{\alpha \rho} \eta^{\beta \sigma} +
\eta^{\alpha \sigma} \eta ^{\beta \rho} ) + {\rm C}( v^{\alpha
}v^{\beta} \eta^{\rho \sigma} + v^{\rho } v^{\sigma} \eta^{\alpha
\beta}) \right. \nonumber \\ & & \left. + {\rm D} v^{\alpha }
v^{\beta} v^{\rho} v^{\sigma} + {\rm E} ( v^{\alpha} v^{\rho}
\eta^{\beta \sigma} +v^{\alpha} v^{\sigma} \eta^{\beta \rho} +
v^{\beta} v^{\rho} \eta^{\alpha \sigma} + v^{\beta} v^{\sigma}
\eta^{\alpha \rho}) \right\} \eeqa We find that each of these
coefficients is independent of the gauge parameter $\alpha$. We also
numerically checked that without the presence of the gauge-fixing term
the propagator could not be inverted.

To compute the preferred-frame effect coefficient $\alpha_2$, we only
need to focus on terms in the momentum-space propagator proportional
to $(v \cdot k)^2$. To leading non-trivial order in $G (v\cdot k)^2$
and in the $c_i$'s we obtain, from the linear combination  $A + 2B +
2C + D + 4E$, \beqa g_{00} &=& 1+ 8 \pi G_{\rm N} \int \frac{d^4 k}{(2
\pi)^4} \frac{1}{k^2} \left\{1 -8\pi G_{\rm N} \frac{(v \cdot
k)^2}{k^2} \frac{1}{c_1(c_1+c_2+c_3)}\left[2 c_1^3 + 4c^2_3(c_2 +c_3)
+ \right. \right. \nonumber \\ & & \left. \left. + c^2_1(3 c_2+5 c_3
+3 c_4) + c_1((6 c_3- c_4)(c_3+c_4) +c_2(6 c_3 +c_4)) \right]
\phantom{\frac{1^2}{1}} \!\!\!\!\!\!\!\! \right\} \tilde{T}^{00}(k)
\eeqa where in the first line $k$ is a four-vector. Next we use
$v^{\mu}=(1,-\vec{w})$, place the source at the origin, substitute
$T^{00} = M \delta ^{(3)}(\vec{x})$ or $\tilde{T}^{00}(k) = 2 \pi M
\delta(k^0)$ and use \beq \int \frac{d^3 k}{(2 \pi)^3} \frac{k_i
k_j}{\vec{k}^4} e^{i \vec{k} \cdot \vec{x}} =\frac{1}{8 \pi r} \left[
\delta_{ij} - \frac{x_i x_j}{r^2} \right] \eeq to obtain \beqa
g_{00}&=& 1 - 2 G_{\rm N} \frac{M}{r} \left(1 - \frac{(\vec{w} \cdot
\vec{r})^2}{r^2} \frac{8 \pi G_{\rm N}}{2c_1(c_1+c_2+c_3)}\left[2
c_1^3 + 4c^2_3(c_2 +c_3) +\right. \right. \nonumber \\ & &
\left. \left. + c^2_1(3 c_2+5 c_3 +3 c_4) +c_1((6 c_3 - c_4)(c_3+c_4)
+ c_2(6 c_3 +c_4)) \right] \phantom{\frac{1^2}{1}} \!\!\!\!\!\!\!\!
\right) \eeqa where we have only written those terms that give a
correction to Newton's law proportional to $[\vec{w}\cdot
\vec{r}/r]^2$. We have also assumed that $|\vec{w}| \ll 1$ so that
higher powers in $\vec{w} \cdot \vec{r}/r$ can be neglected. The
factor of $1/c_1$ in the preferred-frame correction to the metric
arises because when $c_1 \rightarrow 0$ the ``transverse'' components
of $\phi^i$ have no spatial gradient kinetic term. Similarly, the
factor of $1/(c_1+c_2+c_3)$ arises because when $c_1+c_2+c_3
\rightarrow 0$ the ``longitudinal'' component of $\phi^i$ has no
spatial gradient kinetic term. Either of these cases causes a
divergence in the static limit.\footnote{This divergence can of course
be avoided by considering higher-derivative terms in the action for
the Goldstone bosons. This would then give non-relativistic dispersion
relations for these modes (as was the case in \cite{nima}).}

The coefficients $c_i$ redefine Newton's constant measured in solar
system experiments and we find that \beq G_{\rm N } = G \left[1 - 8
\pi G(c_1 + c_4) \right] \simeq \frac{G}{1+ 8 \pi G_{\rm } (c_1 +c_4)}
\eeq which agrees with previous computations to linear order in the
$c_i$'s after correcting for the differences in notation
\cite{carroll,unitvtreview}.

The experimental bounds on deviations from Einstein gravity in the
presence of a source are usually expressed as constraints on the
metric perturbation. Since the metric is not gauge-invariant, these
bounds are meaningful only once a gauge is specified. In the
literature, the bounds are typically quoted in harmonic gauge. Here,
the preferred-frame effect is a particular term appearing in the
solution for $h_{00}$. For static sources, the gauge transformation
needed to translate the solution in our gauge to the harmonic gauge is
itself static. But since a static gauge transformation cannot change
$h_{00}$, we may read off the coefficient of the preferred-frame
effect in the gauge that we used.

By inspection \beqa \alpha_2 &=& \frac{8 \pi G_{\rm
N}}{c_1(c_1+c_2+c_3)}\left[2 c_1^3 + 4c^2_3(c_2 +c_3) +c^2_1(3 c_2+5
c_3 +3 c_4) \right. \nonumber \\ & & \left. +c_1( (6c_3 -c_4)(c_3 +
c_4) + c_2(6 c_3 +c_4)) \phantom{1^2} \!\!\!\!\!\! \right]~, \eeqa
which can be compared with the experimental bound $|\alpha_2| < 4
\times 10^{-7}$ given in \cite{Will2}.

A considerably stronger constraint on the size of the $c_i$'s can
probably be obtained from the gravitational Cherenkov radiation of the
highest-energy cosmic rays \cite{nelson}.

{\bf Acknowledgments}: We thank Ted Jacobson, Kip Thorne, and Matthew
Schwartz for discussions, and Guy Moore for pointing out the error in
the derivation of an observational bound on $\alpha_2$ that appeared
in an early version of this manuscript. This work was supported by the
Department of Energy under the contract DE-FG03-92ER40701.


\end{document}